\documentclass[amsmath,amssymb,aps,pre,superscriptaddress,citeautoscript,floatfix]{revtex4-2}
\usepackage{graphicx}
\usepackage[bookmarks=false]{hyperref}
\usepackage{xcolor}
\usepackage[useregional]{datetime2}

\begin{document}

\title{Theory of inhomogeneous rod-like Coulomb fluids}

\author{Rudolf Podgornik}
\affiliation{School of Physical Sciences and Kavli Institute for Theoretical Sciences, University of Chinese Academy of Sciences, Beijing 100049, China} 
\affiliation{Wenzhou Institute of the University of Chinese Academy of Sciences, Wenzhou, Zhejiang 325000, China}
\affiliation{Institute of Physics, Chinese Academy of Sciences, Beijing 100190, China}
\email{podgornikrudolf@ucas.ac.cn Also affiliated with Department of Physics, Faculty of Mathematics and Physics, University of Ljubljana, SI-1000 Ljubljana, Slovenia and Department of Theoretical Physics, Jo\v zef Stefan Institute, SI-1000 Ljubljana, Slovenia.}

\begin{abstract}
A field theoretic representation of the classical partition function is derived for a system composed of a mixture of anisotropic and isotropic mobile charges that interact {\sl via} long range Coulomb and short range nematic interactions. The field theory is then solved on a saddle-point approximation level, leading to a coupled system of  Poisson-Boltzmann and Maier-Saupe equations. Explicit solutions are finally obtained for a rod-like counterion-only system in proximity of a charged planar wall, generalizing the standard Gouy-Chapman results. The nematic order parameter profile, the counterion density profile and the electrostatic potential profile are interpreted within the framework of a nematic wetting layer with a (Donnan) potential difference. 
\end{abstract}

\maketitle

\section{Introduction}

Coulomb fluids composed of anisotropic charge carriers are ubiquitous in many contexts. In fact it is worth noting that strong electrostatic interactions between rod-like charges were already invoked in the case of nematic ordering of tobacco mosaic virus (TMV) in the seminal work of Bernal and Fankuchen \cite{Bernal}, which is also one of the first cases of the application of the Poisson-Boltzmann theory to biological systems  \cite{Review1}.  Apart from the TMV, other charged rod-like viruses and virus-like nanoparticles have been used in functional materials assembly \cite{assembly2} whose formation is controlled by the electrostatic interaction with the substrate  \cite{assembly4}. 

A strong electrostatic attraction between the substrate and the deposited filamentous viruses enhances a stable film formation, as is clear from the effect of the ionic strength and the $p$H of the solution \cite{assembly3,assembly1}. Different types of filamentous viruses \cite{Tang1996}, as well as short fragments of DNA \cite{Bellini2012}, F-actin  \cite{Sanders15994}, and cellular scaffold microtubules \cite{Needleman16099} all exhibit also properties of polyvalent rod-like ions as do also many other multivalent strongly anisotropic biological polyvalent ions \cite{Teif2011} that can be either modeled as spatially distributed point charges or as higher order point multipoles \cite{Kanduc2009, Bohinc2015, Naji2018}. Last but not least, ionic liquid crystals (ILCs) \cite{Fernandez2016} are solvent-less ionic systems with a dual ionic and organic nature \cite{RuiShi2016}, that are composed of cations and anions with at least one ionic species and characterized by highly anisotropic molecular shape \cite{Goossens2016}. 

The rod-like shape of ions leads to ordered structures whose formation exhibits features of liquid crystal ordering as well as long-range electrostatic interactions.  It is this latter example that has recently witnessed a real proliferation of theoretical works set to illuminate the basic principles of order formation in these complex Coulomb systems.

The theoretical approaches to charged anisotropic systems in the bulk are varied and abundant. Here we delimit to a few that are directly relevant for subsequent developments.  The work of Deutsch and Goldenfeld \cite{deutsch, Deutsch82} for thin charged rods relies on a collective coordinate transformation method applied to the ordering of charged rods in 3D. In addition, for rod-like charged cylinders a generalized Onsager theory could also be used to describe the ordering transition with electrostatic interactions strongly modifying the hard core diameter of the rods as well as providing a mechanism for twisting interaction as first described by Odijk \cite{Odijk86,Schoot1990,Vroege1992}. This approach has seen many further developments with different level modifications and extensions ~\cite{Trizac14, Jho14,PhysRevE.63.061705,PhysRevE.79.041401, doi:10.1063/1.1739393}. A generalized variational field theory of particles with rigid charge distributions ~\cite{LUE2006236} and an order parameter based mean-field approximation of rod-like polyelectrolytes~\cite{Muthu99} both lead to an ordering transition in 3D. 

The properties of bulk systems composed of a mixture of multipolar charges have also been analyzed in detail based on a field theoretic approach that naturally incorporates also non-local dielectric response \cite{Andelman2013,Sahin2,Frydel2016}, while the nematic order and electrostatics in the case of ion-doped nematic electrolytes, with an anisotropic dielectric response, have been considered on a phenomenological level ~\cite{Ravnik2020}.  Bulk properties as well as electrical double layers in  ionic liquid crystals have been analyzed in the work of Bier within the density functional approach  \cite{Bier2007, Kondrat2010} that was formulated for homogeneous as well as inhomogeneous systems with interfaces \cite{Bier2005,Bier2006}. It is the latter case that is particularly interesting as it should exhibit features of both, the nematic ordering as well as the Gouy-Chapman-type electrostatic double layers. 

\begin{figure*}[t!]
\centering
\includegraphics[width=.8\textwidth]{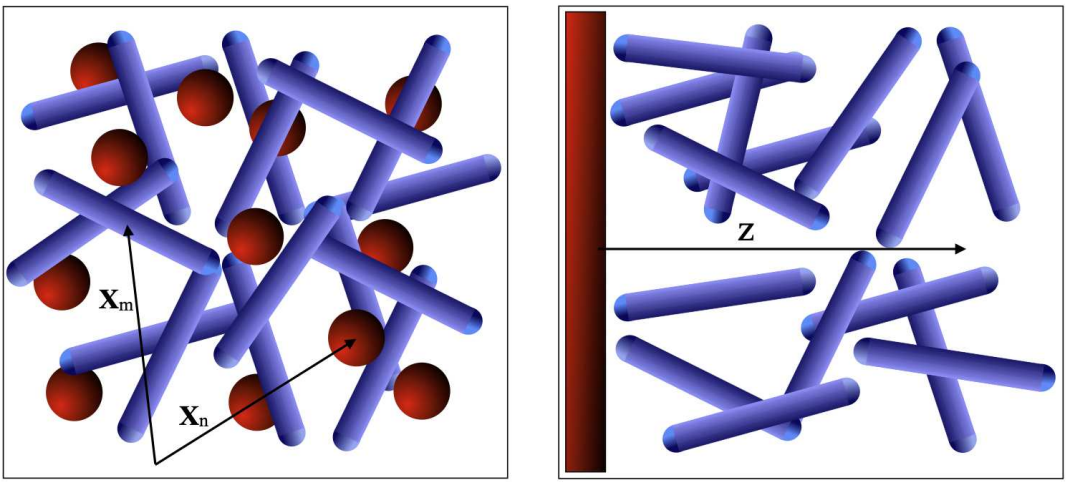}
\caption{A schematic presentation of the system with rod-like cations and simple anions in the bulk (left). An inhomogeneous system of rod-like cations close to a charge surface (right). Both the cations as well as the anions are charged, though in general differently. Apart from long range Coulomb interactions the rod-like cations interact also via short range nematic interactions described by a Lebwohl-Lasher interaction potential. \label{graph0}}
\end{figure*} 

The problem of an inhomogeneous system with multipolar anisotropic charges has also been addressed  on various levels of modelling and approximations.  %, rather then the general problem of rod-like Coulomb fluids, has not been analyzed to the same degree as the bulk systems but this is exactly what we are interested in here.  The  inhomogeneous case 
The case of a system bounded by charged wall(s) with mobile dipolar charges has been analyzed with field theoretical approach \cite{Abrashkin2007, AZUARA20085587,Sahin1}, while quadrupolar charges \cite{Kanduc2009, doi:10.1063/1.4865878} and finite size dumbbell charges \cite{PhysRevLett.101.208305, Bohinc2015} have been investigated on different levels of approximations, including field theoretical approach. Density functional theory was formulated for the case of inhomogeneous systems of charged anisotropic particles with interfaces  \cite{Bier2005, Bier2006, Bier2007}, specifically a semiinfinite isotropic or nematic bulk system in contact with a charged hard wall exhibiting nematic wetting of the substrate, which is close to our point of departure. 

Our focus here will be on how to formulate a theory of an inhomogeneous system of anisotropic charges with  {\sl Coulomb and nematic interactions} - as in the case of rod-like cations close to an oppositely charged wall - based on a formalism that could be seen as a {\sl straightforward generalization of the Gouy-Chapman theory} for simple ions. Thus, in what follows we will derive a statistical theory of a Coulomb system composed of anisotropic rod-like cations and point-like anions with microscopic interactions of the simple Lebwohl-Lasher nematic as well as Coulomb type and derive its mean-field approximation form.  We will then apply the general theory to the case of an inhomogeneous system, composed of a negatively charged planar interface in the presence of rod-like counterions, thus generalizing the standard counterion-only Gouy-Chapman problem of colloid physics. We will derive the equations governing the density distribution and electrostatic potential on the mean-field level, in the case of a one dimensional system with an electrified interface. By solving the mean-field level equations that emerge as a coupled system of the Maier-Saupe and the Poisson-Boltzmann equation, we are able to derive some salient properties of inhomogeneous nematic ordering induced by the charged interface as well as the modifications in the electric double layer distribution wrought by the presence of nematic order. 

\section{Collective description and field theoretic representation}
\subsection{Order parameters}

Let us consider a system composed of monovalent anions (charge $-e$) and polyvalent rod-like cations (charge $+ qe$) with  chemical potential $\mu$, a situation often encountered in many ionic liquid system with cations being extended stiff rods, while anions are considered to be point-like particles, see Fig. \ref{graph0}. 

The respective microscopic order parameters are the {\sl nematic order parameter density} defined as (see {\sl e.g.} Ref. \cite{Frenkel1991})
\begin{eqnarray}
\hat{\cal Q}_{ij}({\bf x}) &\equiv&  \sum_{(+)} {\textstyle\frac32} \left( {\bf n}_i {\bf n}_j - {\textstyle\frac13} \delta_{ij}\right)\delta\left( {\bf x} - {\bf x}_n \right),
\label{Qdef1}
\end{eqnarray}
with ${\bf n}({\bf x}_n)$ the director of the $n-th$ rod-like cation, together with the microscopic {\sl charge density}
\begin{eqnarray}
&&\hat{\rho}_{(\pm)}({\bf x}) \equiv  (+q, -1) \sum_{(\pm)} \delta\left( {\bf x} - {\bf x}_n \right)\label{Qdef2}
\end{eqnarray}
so that the total charge density is $ \hat{\rho}({\bf x}) = \hat{\rho}_{(+)}({\bf x}) + \hat{\rho}_{(-)}({\bf x})$. The cations are polyvalent, with valency $q$, and the anions are univalent $q=1$. The ensemble averaged forms of the above order parameters are defined as
\begin{eqnarray}
\Big<\hat{\cal Q}_{ij}({\bf x})\Big> = c_+({\bf x}) ~{\cal S}({\bf x}) ~{\textstyle\frac32} \left( {\bf n}_i({\bf x}) {\bf n}_j({\bf x}) - {\textstyle\frac13} \delta_{ij}\right)\label{defins}
\end{eqnarray}
and 
\begin{eqnarray}
 \Big<\hat{\rho}_{+}({\bf x})\Big>  = q c_{+}({\bf x}), \qquad \Big<\hat{\rho}_{-}({\bf x})\Big>  = - c_{-}({\bf x}),
\end{eqnarray}
where the corresponding thermodynamic densities of the rod cations and anions are $ c_+({\bf x}), c_{-}({\bf x})$ and the nematic scalar order parameter is ${\cal S}({\bf x}) $.

The microscopic Hamiltonian is assumed to be of the general Lebwohl-Lasher type, though in the original formulation it was assumed to act only between nearest neighbors \cite{Lebwohl1972}. This constraint was relaxed in a recent analytical formulation \cite{BingSuiLu}. By assumption then
\begin{eqnarray}
{\cal H}[{\bf r}_n, {\bf n}_n ] &=&  {\textstyle\frac12} \sum_{n,m} ~u_{QQ}({\bf x}_n -{\bf x}_m) \left(\Big({\bf n}_n{\bf n}_m\Big)^2 - {\textstyle\frac13} \right)  + \nonumber\\
&& + {\textstyle\frac12}  \sum_{n,m} ~u_{\rho\rho}({\bf x}_n -{\bf x}_m),
\label{hamil1}
\end{eqnarray}
where $n, m$ run over all the particles in the system and ${\bf n}_n,~{\bf n}_m$ are the unit directors of the $n$-th and $m$-th particles. The interaction strengths are all expressed in thermal units. The Lebwohl-Lasher interaction type is chosen in order to simplify the derivation but is not essential and can be easily generalized. The Onsager interactions are at least to the lowest order equivalent to the first term in Eq. \ref{hamil1}, where the nematic interaction has a delta function range \cite{Doi1981}. The theory presented here therefore incorporates to the lowest order also the standard Onsager results. 

The interaction energy Eq. \ref{hamil1} can be clearly recast as
\begin{eqnarray}
{\cal H}[{\bf r}_n, {\bf n}_n ] &=& {\textstyle\frac12} \int\!\!\int_V d{\bf x} d{\bf x}' ~\hat{\cal Q}_{ij}({\bf x}) u_{QQ}({\bf x} -{\bf x}') \hat{\cal Q}_{ij}({\bf x}') + \nonumber\\
&& + {\textstyle\frac12} \int\!\!\int_V d{\bf x} d{\bf x}' ~\hat{\rho}({\bf x})u_{\rho\rho}({\bf x} -{\bf x}') \hat{\rho}({\bf x}'),
\label{hamilt0}
\end{eqnarray}
where we further assume that the scalar part $u_{\rho\rho}({\bf x} -{\bf x}')$ is due to the long-range Coulomb interactions. The coupling between the two is omitted to the lowest order but can be considered for e.g. dipolar or quadrupolar rods.  The self-energies which were also omitted from the formula above can be absorbed into the chemical potential when defining the grand canonical partition function.

\subsection{Field theoretical representation and thermodynamic relations}

In the Appendix we derive the field theoretical representation of the grand canonical partition function, $\Xi$,  in terms of the tensor, $\Phi_{ij}({\bf x})$, and scalar, $\phi({\bf x})$, auxiliary fields with
\begin{eqnarray}
{\Xi}[\Phi_{ij}({\bf x});  \phi({\bf x})]    &=& \int\!\!{\cal D}[\Phi_{ij}({\bf x})] \int\!\!{\cal D}[\phi({\bf x})]~ e^{-\beta{\underline{\cal H}}[\Phi_{ij}({\bf x});  \phi({\bf x})] }, 
\label{cgkfajbf}
\end{eqnarray}
with the effective field action as
\begin{widetext}
\begin{eqnarray}
-\beta {\underline{\cal H}}[ \Phi_{ij}({\bf x}); ~\phi({\bf x})] &=& ~-{\textstyle\frac12} \int\!\!\int_V d{\bf x} d{\bf x}' ~ {\Phi}_{ij}({\bf x}) u_{QQ}^{-1}({\bf x} -{\bf x}')  {\Phi}_{ij}({\bf x}') - {\textstyle\frac12} \int\!\!\int_V d{\bf x} d{\bf x}' ~ {\phi}({\bf x})u_{\rho\rho}^{-1}({\bf x} -{\bf x}') {\phi}({\bf x}') + \nonumber\\
&&  + ~\lambda_{(+)} \int_V d{\bf x} ~e^{ i\phi({\bf x}) +\ln{{\cal P}(\Phi_{ij}({\bf x}))}}+  \lambda_{(-)} \int_V d{\bf x} ~e^{-i\phi({\bf x})},
\label{fieldtyui}
\end{eqnarray}
\end{widetext}
where $\lambda_{(\pm )}$ are the fugacities of the two ionic species and  ${\cal P}(\Phi_{ij}({\bf x}))$ is the orientational partition function of a single particle, see Eq. \ref{Maier}, 
\begin{eqnarray}
{\cal P}(\Phi_{ij}({\bf x})) \equiv \Big< e^{i {\textstyle\frac32}\left({\bf n}_i {\bf n}_j - {\textstyle\frac13}\delta_{ij} \right) \Phi_{ij}({\bf x})} \Big>_{\Omega}.
\label{distrib}
\end{eqnarray}
The orientational trace is defined in such a way that $<1>_{\Omega} = 1$.  From the expressions above it is clear that for the cations the interaction with the fields has a scalar, electrostatic component, and a tensor nematic component given by the orientational entropy. While the field theoretical representation of the partition function cannot be solved analytically, it suggests a number of developments that lead to meaningful approximations, such as the saddle-point approximation that we will detail in what follows. In addition other approximations that turned out to be useful in the context of isotropic Coulomb fluids could be exploited as well \cite{Coulomb}.

There is a number of  thermodynamic relations that need to be satisfied by the derivatives of the partition function. In the case of a set chemical potential, the fugacity and the average number of molecules in the system are given by the standard relations. In addition, the invariance of the functional integral with respect to the linear transformation of the fluctuating fields yields two relations
\begin{eqnarray}
\Big< \frac{\delta {\underline{\cal H}}[\Phi_{ij}({\bf x}); ~\phi({\bf x})]}{\delta \Phi_{ij}({\bf x})}\Big> = 0
\qquad {\rm and} \qquad 
\Big< \frac{\delta {\underline{\cal H}}[\Phi_{ij}({\bf x}); ~\phi({\bf x})]}{\delta \phi({\bf x})}\Big> = 0,
\label{SD2}
\end{eqnarray} 
where the averages stand for 
\begin{eqnarray}
\!\!\!\!\!\Big< \dots \Big> = \frac{\int {\cal D}[\Phi_{ij}({\bf x})] \int {\cal D}[\phi({\bf x})]~ \dots  ~e^{-\beta{\underline{\cal H}}[\Phi_{ij}({\bf x}); \phi({\bf x})] }}{\int {\cal D}[\Phi_{ij}({\bf x})] \int {\cal D}[\phi({\bf x})]~ e^{-\beta{\underline{\cal H}}[\Phi_{ij}({\bf x}); \phi({\bf x})] }}.~~~~
\end{eqnarray} 
These two relations, Eq. \ref{SD2}, furthermore yield the following exact connections between the average values of the order parameters and the auxiliary fields
\begin{eqnarray}
&& \Big< \hat{\cal Q}_{ij}({\bf x})\Big> = - i \int_V d{\bf x}'~ u^{-1}_{QQ}({\bf x} -{\bf x}') \Big<\Phi_{ij}({\bf x}')\Big> = \nonumber\\
&& = -  i \lambda_{(+)}~ \Big<\frac{\partial \ln{{\cal P}(\Phi_{ij}({\bf x}))}}{\partial \Phi_{ij}({\bf x})} e^{ i\phi({\bf x})+ \ln{{\cal P}(\Phi_{ij}({\bf x}))}}\Big> = \nonumber\\
&& = \lambda_{(+)}~ \Big< \Big<\!\!\!\Big<{\textstyle\frac32}\left({\bf n}_i {\bf n}_j - {\textstyle\frac13}\delta_{ij} \right) \Big>\!\!\!\Big>_{\Omega}  e^{ i\phi({\bf x}) + \ln{{\cal P}(\Phi_{ij}({\bf x}))}}\Big>, 
\label{equival1}
\end{eqnarray}
where we took into account the definition of the double brackets Eq. \ref{doublebrac}. Furthermore by analogy
\begin{eqnarray}
&& \Big< {\rho}({\bf x})\Big> = - i \int_V d{\bf x}'~ u^{-1}_{\rho\rho}({\bf x} -{\bf x}') \Big<\phi({\bf x}') \Big> = \nonumber\\
&& ~~~~+ q \lambda_{(+)}  \Big<~e^{ i\phi({\bf x}) +\ln{{\cal P}(\Phi_{ij}({\bf x}))}}\Big> - \lambda_{(-)} ~\Big< e^{-i\phi({\bf x})}\Big>. 
\label{equival2}
\end{eqnarray}  
Note the difference between the orientational averages $ \Big<\!\!\!\Big< ... \Big>\!\!\!\Big>_{\Omega}$ ,  $\Big< \dots \Big>_\Omega$ defined as
\begin{eqnarray}
\Big<\!\!\!\Big< ... \Big>\!\!\!\Big>_{\Omega} = \frac{\Big< (\dots)e^{i {\textstyle\frac32}\left({\bf n}_i {\bf n}_j - {\textstyle\frac13}\delta_{ij} \right) \Phi_{ij}({\bf x})} \Big>_{\Omega}}{\Big< e^{i {\textstyle\frac32}\left({\bf n}_i {\bf n}_j - {\textstyle\frac13}\delta_{ij} \right) \Phi_{ij}({\bf x})} \Big>_{\Omega}}
\label{doublebrac}
\end{eqnarray} 
where the unnormalized orientational average, $\Big< \dots \Big>_\Omega$, is with respect to the distribution Eq. \ref{distrib}. 

On the saddle-point  level we will soon see that the above two equations are actually the modified Maier-Saupe (MS)  self-consistent equation and the modified Poisson-Boltzmann (PB) self-consistent equation for the tensor and scalar fields, respectively.

\subsection{Saddle-point approximation}

Since the field action Eq. \ref{fieldtyui}  is non-linear, no further exact developments are feasible and one needs to resort to the {\sl saddle-point approximation} that yields two mean-field equations for the two auxiliary fields.  At the saddle-point the fields are pure imaginary, so that one can transform 
\begin{equation}
\Phi_{ij}({\bf x}) \longrightarrow - i \Phi_{ij}^*({\bf x}) \quad {\rm and} \quad \phi({\bf x}) \longrightarrow i \phi^*({\bf x}).
\label{hdgkryn}
\end{equation}
The thermodynamic averages $\Big< \dots \Big>$ are given at the value of the mean-field, and the two self-consistent field equations, Eqs. \ref{equival1}, \ref{equival2}, are then reduced to
\begin{eqnarray}
&&\Big< \hat{\cal Q}_{ij}({\bf x})\Big>  =  \lambda_{(+)}~ \Big<\!\!\!\Big<{\textstyle\frac32}\left({\bf n}_i {\bf n}_j - {\textstyle\frac13}\delta_{ij} \right) \Big>\!\!\!\Big>_{\Omega}  e^{ -\phi^*({\bf x})+ \ln{{\cal P}(-i\Phi^*_{ij}({\bf x}))}} = c_{(+)}(\phi^*({\bf x}))  \Big<\!\!\!\Big<{\textstyle\frac32}\left({\bf n}_i {\bf n}_j - {\textstyle\frac13}\delta_{ij} \right) \Big>\!\!\!\Big>_{\Omega}, 
\label{conne}
\end{eqnarray}
with the orientational partition function ${\cal P}(-i\Phi^*_{ij}({\bf x}))$ defined in Eq. \ref{distrib} taken at the imaginary value of the tensor auxiliary field, and
\begin{eqnarray}
&& \Big< {\rho}({\bf x})\Big> =  + q \lambda_{(+)} ~e^{ -\phi^*({\bf x}) +\ln{{\cal P}(-i\Phi^*_{ij}({\bf x}))}}  - \lambda_{(-)} ~ e^{+\phi^*({\bf x})}= + q c_{(+)}(\phi^*({\bf x})) - c_{(-)}(\phi^*({\bf x})), 
 \end{eqnarray}  
with obvious definitions of the cation and anion densities, $c_{(+)}(\phi^*({\bf x}))$ and $c_{(-)}(\phi^*({\bf x}))$.
In what follows we will assume a short range attractive orientational potential and a Coulombic positional potential. This implies
\begin{eqnarray}
u_{QQ}^{-1}({\bf x} -{\bf x}') = - u_{QQ}^{-1}(0)\delta({\bf x} -{\bf x}')
\label{coulombic}
\end{eqnarray}
and
\begin{eqnarray}
u_{\rho\rho}^{-1}({\bf x} -{\bf x}') = - \varepsilon\nabla^2 \delta({\bf x} -{\bf x}')
\label{xacb}
\end{eqnarray}
where $\varepsilon \equiv \epsilon_0\epsilon$, with $\epsilon$ the dielectric permittivity of the solvent and $\epsilon_0$ the permittivity of space. Other forms of the nematic interaction potential are of course possible but would lead to more complicated tensorial saddle-point equations. 

With this in mind we then derive the tensorial part in the form of a modified Meier-Saupe (MS) equation
\begin{eqnarray}
\!\!\!\!\!\!{- u_{QQ}^{-1}(0) ~\Phi^*_{ij}({\bf x}) + c_{(+)}~  \Big<\!\!\!\Big<{\textstyle\frac32}\left({\bf n}_i {\bf n}_j - {\textstyle\frac13}\delta_{ij} \right) \Big>\!\!\!\Big>_{\Omega} = 0}.
\label{MSgui1}
\end{eqnarray}
We will see that the Maier-Saupe equation determines only the nematic order parameter but not the orientation in the ordered phase, which is assumed to be homogeneous. The orientation of the ordering would be determined by the substrate-nematic interaction potential in line with the model described in \cite{Braun_1996}. Without any loss of generality we can assume the ordering is perpendicular to the bounding surface. To describe the orientational relaxation effects one would need also the elastic deformation energy which would stem from the expansion of the nematic interaction potential w.r.t. the gradient of the tensorial order parameter as in the general inhomogeneous Landau-de Gennes {\sl Ansatz} \cite{Braun_1996}. We assume that the that the characteristic length of the nematic order relaxation is much shorter then the electrostatic relaxation length, either the Gouy-Chapman length or the Debye length, and can thus be neglected. We discuss the possible generalizations in the Discussion section.

The scalar part of the mean-field equations can be written in the form of the Poisson-Boltzmann (PB) equation as
\begin{eqnarray}
{\varepsilon \nabla^2\phi^*({\bf x})  + q c_{(+)} - c_{(-)} = 0}.
\label{MF2}
 \end{eqnarray}  
The two mean-field equations Eqs. \ref{MSgui1} and \ref{MF2} correspond to the nematic and electrostatic degrees of freedom in a similar manner that the Edwards and the polymer Poisson-Boltzmann equation correspond to polymer and electrostatic degrees of freedom for charged flexible polymers \cite{Podgornik1992,Borukhov_1995}. Consequently any other degree of freedom would introduce its own mean-field equation. Clearly the approximations entailed in the derivation of the mean-field equations imply that the spatial relaxation of both fields is given solely by the electrostatic component, so that in this respect the nematic field is subservient to the electrostatic field.

Inserting the mean-field {\sl Ansatz} into the free energy Eq. \ref{fieldtyui} we obtain the general form of the mean-field free energy ${\cal F}[\Phi_{ij}^*({\bf x}); ~\phi^*({\bf x})] $  as a functional of the {\sl mean-field nematic potential}, $\Phi_{ij}^*({\bf x})$,  and the {\sl mean-field electrostatic potential}, $\phi^*({\bf x})$  as 
\begin{widetext}
\begin{eqnarray}
\beta {\cal F}[\Phi_{ij}^*({\bf x}) ; ~\phi^*({\bf x})] =  &=& ~-{\textstyle\frac12} \int\!\!\int_V d{\bf x} d{\bf x}' ~ \Phi_{ij}^*({\bf x}) u_{QQ}^{-1}({\bf x} -{\bf x}')  \Phi_{ij}^*({\bf x}') - {\textstyle\frac12} \int\!\!\int_V d{\bf x} d{\bf x}' ~ \phi^*({\bf x})u_{\rho\rho}^{-1}({\bf x} -{\bf x}') \phi^*({\bf x}') - \nonumber\\
&&  - ~\lambda_{(+)} \int_V d{\bf x} ~e^{ -\phi^*({\bf x}) +\ln{{\cal P}(-i\Phi_{ij}({\bf x}))}} - \lambda_{(-)} \int_V d{\bf x} ~e^{+\phi^*({\bf x})}.
\label{mjtybjo}
\end{eqnarray}
\end{widetext}
The mean-field theory for rod-like cations thus couples the MS equation for simple liquid crystals to the PB equation for simple Coulomb fluid, except that formally the electrostatic potential of the rod-like cations is transformed to $\phi^*({\bf x})  \longrightarrow \phi^*({\bf x}) - \ln{{\cal P}(-i\Phi^*_{ij}({\bf x}))}$. 
Clearly one could expand the above free energy in terms of the nematic order parameter obtaining a generalized Landau-de Gennes free energy but then lose the direct connection with the Gouy-Chapman theory. As already stated in the Introduction we aim to keep and develop this connection.

The theory was formulated specifically for a rod-like cationic species and a point-like anionic species, but based on the methodology other possibilities are just as amenable to the same procedure of deriving the field theory as well as the mean-field equations.

\section{Rod-like counterion-only system in one dimension}

There are of course countless cases that one can dwell on in order to apply the general theory, and thus  some selectivity is in order. In what follows we shall delimit ourselves to the case of a rod-like counterion-only system in the presence of a single electrified surface, which is a direct generalization of the standard Gouy-Chapman paradigm of the colloid electrostatics. The only spatial dependence is in the direction of the surface normal, which we choose to coincide with the $z$ axis. One also needs to remember that the counterion only system has no reservoir and thus no chemical potential, so that the number of counterions is set only by the neutralizing charge on the bounding surface. Nevertheless this does not affect the above derivation.

\subsection{Coupled system of Maier-Saupe and Poisson-Boltzmann equations}

The scalar mean-field equation in this case  is the modified Poisson-Boltzmann equation that is obtained as
\begin{eqnarray}
&& \varepsilon {\phi^*}''(z) + q c_{(+)}(\phi^*(z) ) = 
0, 
\label{MF22}
\end{eqnarray}  
where ${\phi^*}''(z) = \frac{d^2 {\phi^*}(z)}{dz^2} $ is the second derivative of the mean potential with respect to $z$. Comparing the above expression with the standard PB equation for a point-like counterion-only case, the difference is manifest in the presence of the orientational entropy of the rods, equal to $\ln{{\cal P}(-i\Phi^*_{ij}({\bf x}))}$. We will solve this equation later. 

The tensor mean-field equation, {\sl i.e.},  the modified Maier-Saupe equation, can be solved by assuming a non-vanishing orientational order characterized with the director $\hat{\bf n}$   
\begin{eqnarray}
\Phi^*_{ij}(z) &=& {s}(z)~ {\textstyle\frac32} \left( \hat{\bf n}_i \hat{\bf n}_j - {\textstyle\frac13} \delta_{ij}\right),
\label{qwergjk}
\end{eqnarray}
so that
\begin{eqnarray}
\Phi^*_{ij}(z)^2 =  {\textstyle\frac32} ~{s}^2(z),
\end{eqnarray}
and ${s}(z)$, proportional but not equal to the nematic order parameter, then simply represents the absolute value of $\Phi^*_{ij}(z)$. In infinite space without any anchoring fields, $ \hat{\bf n}$ is arbitrary but would be set by surface free energy terms \cite{Braun_1996}, which we did not invoke explicitly since the Maier-Saupe equation is local and thus valid for any orientation.   One can recall, Eq. \ref{defins} and Eq. \ref{qwergjk}, which means that
\begin{eqnarray}
&& {\cal Q}^*_{ij}  =  u^{-1}_{QQ}(0) \Phi^*_{ij}  =  u^{-1}_{QQ}(0) {s} ~ {\textstyle\frac32} \left( \hat{\bf n}_i \hat{\bf n}_j - {\textstyle\frac13} \delta_{ij}\right) = c_+  ~{\cal S}  ~{\textstyle\frac32} \left( \hat{\bf n}_i  \hat{\bf n}_j  - {\textstyle\frac13} \delta_{ij}\right) 
\label{qwergjk1}
\end{eqnarray}
wherefrom we derive the exact connection between the nematic order parameter ${\cal S}$, strictly limited to the interval $0 \leq {\cal S} \leq 1$,  and the parameter $s$ introduced in Eq. \ref{qwergjk} as
\begin{equation}
{\cal S} = \frac{ s}{u_{QQ}(0)~c_+}.
\label{vcfs}
\end{equation}
By multiplying both sides of Eq.\ref{MSgui1} by $ \left( \hat{\bf n}_i \hat{\bf n}_j - {\textstyle\frac13} \delta_{ij}\right)$ and tracing over the indices, we finally obtain
\begin{eqnarray}
{s}(z) &=& {\textstyle\frac32}  u_{QQ}(0) %\lambda_{(+)} e^{ -\phi^*(z)+ \ln{{\cal P}(-i\Phi^*_{ij}(z))}}
c_{(+)}(\phi^*(z))~\Big<\!\!\!\Big< ({\bf n} \cdot \hat{\bf n})^2 - {\textstyle\frac13} \Big>\!\!\!\Big>_{\Omega}.
\label{MSgui23}
\end{eqnarray}
Since the above relation  is a local one, it remains the same at any position $z$, and the explicit dependence on the coordinate can be dropped in what follows. This is a consequence of the assumption that the nematic order relaxation is much shorter then the electrostatic relaxation length. Developing further, this eventually yields
\begin{eqnarray}
\!\!\!\!&&\Big<\!\!\!\Big< ({\bf n} \cdot \hat{\bf n})^2 - {\textstyle\frac13} \Big>\!\!\!\Big>_{\Omega} = \Big<\!\!\!\Big< \cos{\theta}^2 - {\textstyle\frac13} \Big>\!\!\!\Big>_{\Omega} =  \frac{\partial }{\partial \gamma} \ln{J(\gamma)},
\label{cadwq}
\end{eqnarray}
where obviously $\gamma =  {\textstyle\frac94} {s}$ and from  Eq. \ref{doublebrac} it furthermore follows that the orientational partition function is obtained as
\begin{eqnarray}
J(\gamma ) = \Big<  e^{\gamma \left(  \cos{\theta}^2 - {\textstyle\frac13} \right)} \Big>_{\Omega} = \int_0^1 dz ~e^{\gamma \left( z^2 -{\textstyle\frac13}\right)}
 =\frac{e^{{\textstyle\frac23} \gamma}}{\sqrt{\gamma}} D(\sqrt{\gamma}), %= \frac{\partial }{\partial \gamma} \ln{J(\gamma)} ,
\end{eqnarray}
with $D(x)$ as the Dawson's integral \cite{Kleinert}. The Maier-Saupe expression can then be cast definitively as 
\begin{eqnarray}
{\gamma} =   
{\textstyle\frac{27}{8}}   u_{QQ}(0) c_+ \frac{\partial }{\partial \gamma} \ln{J(\gamma)}
\label{finalS}
\end{eqnarray}
where the nematic order parameter is then extracted from Eq. \ref{vcfs} as
${{\cal S} = {\gamma}/({\frac{9}{4} u_{QQ}(0)~c_+}}).$ In 2D the same type of analysis would just replace the Dawson integral by a modified Bessel function integral \cite{sunita2020}. Eq. \ref{finalS} corresponds exactly to the Kleinert formulation \cite{Kleinert} of the Maier-Saupe theory, if one takes his coupling strength of nematic interactions equal to ${\cal A}_0 = {\textstyle\frac32} u_{QQ}(0) c_+$. 

Note again that the mean-field Maier-Saupe equation is a local equation that pertains to every point in the domain, which is a consequence of the fact that in the {\sl Ansatz} Eq. \ref{coulombic} we only considered local nematic interactions. 

The solution of the Maier-Saupe  equation for $0 \leq {\cal S} \leq 1$ exhibits a first order isotropic-nematic transition at a critical value of $c_{(+)} $,  where the order parameter makes a discontinuous jump from an isotropic phase to a nematic phase. It is important to realize that the solutions of the MS theory for this counterion only system differ from the standard case, where the two phases are kept under the same chemical potential that allows for the density jump at the transition. The counterion only system is not regulated by a chemical potential and exhibits no density jump.

\subsection{First integral and the phase portrait analysis}

\begin{figure*}[t!]
\centering
\includegraphics[width=.8\textwidth]{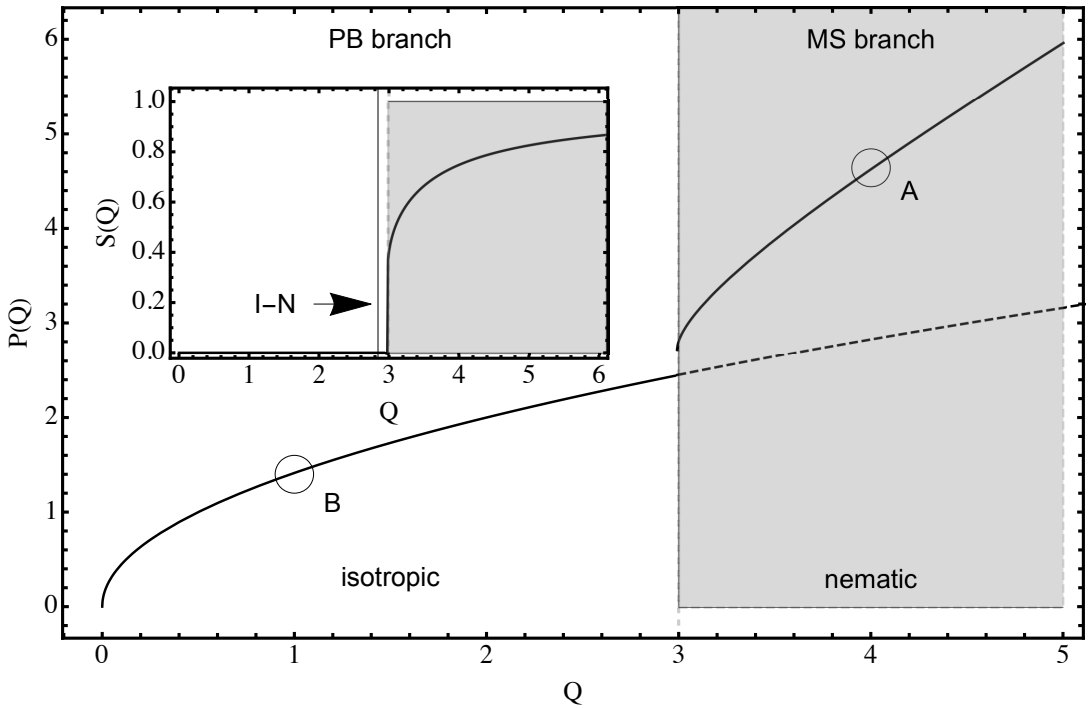}
\caption{Phase portrait solution of the Maier-Saupe equation. Phase portrait $P(Q)$ obtained from the mean-field equation, Eq. \ref{Hamilt4}. The solution starts with $P_0$ obtained from the boundary condition Eq. \ref{Hamilt5} and then moves towards $P = 0$ far away from the surface. If $P_0$ is above the IN transition point, e.g. "A", the solution first follows the MS branch (full line) and then at the transition migrates to the PB branch that it follows until the electrostatic field levels off to zero far away from the surface, just as in the case of point-like counterions. On the other hand, if $P_0$ is below the transition point, e.g. "B", the solution simply follows the PB line as if the counterions are point-like. (Inset) The dependence of the nematic order parameter ${\cal S}$ on the dimensionless density $Q$, defined in Eq. \ref{bdfka}. The isotropic-nematic (IN) transition takes place at a critical value $Q_c \simeq 3.$ and the jump in the order parameter amounts to $0.313$.   \label{graph1}}
\end{figure*} 

Introducing now the generalized van't Hoff osmotic pressure as
\begin{eqnarray}
p(\phi^*(z), s(z)) &=& 
c_{+}(\phi^*(z),  s(z)),
\label{osmot}
\end{eqnarray}
it is possible to write the mean-field equations in a "Lagrangian" form (see Ref. \cite{C5SM01757B})
\begin{eqnarray}
& & \varepsilon \phi^*(z)''  = \frac{\partial p(\phi^*(z), s(z))}{\partial \phi^*(z)}, \nonumber\\
& & s(z)  =  - {\textstyle\frac23}  u_{QQ}(0)  \frac{\partial p(\phi^*(z), s(z))}{\partial s(z)}.
\label{mhpy}
\end{eqnarray}
Proceeding now as in the case of the first integral of the Gouy-Chapman theory \cite{Safinya}, by multiplying the first equation by ${\phi^*}' $ and the second one by $s'(z) $ and then summing them up, we obviously remain with
\begin{widetext}
\begin{equation}
\frac{\partial p}{\partial \phi^*} {\phi^*}' + \frac{\partial p}{\partial  s(z)} s'(z) = \Big( \varepsilon \phi^*(z)'' \phi^*(z)' +  {\textstyle\frac32}  u_{QQ}(0)^{-1} s(z) s'(z) \Big) = \frac{d}{dz}\Big( {\textstyle\frac12}\varepsilon  ({\phi^*}'(z))^2 - {\textstyle\frac34} ~ {u_{QQ}(0)}^{-1}s^2(z)\Big),
\label{bcjsgb}
\end{equation}
\end{widetext}
which can be cast into the form of the first integral that generalizes the standard Poisson-Boltzmann result 
\begin{eqnarray}
\!\!\!\!\!\!\!{\textstyle\frac12}\varepsilon  ({\phi^*}'(z))^2 - {\textstyle\frac34} \frac{s^2(z)}{{u_{QQ}(0)}} - p(\phi^*(z), s(z)) = const. \label{ncfgjkws}
\end{eqnarray}
Because of the assumption of the short range nematic interactions, the field $s(z)$ has no associated "dynamics", {\sl i.e.}, the first integral contains no derivatives of $s(z)$. 

The "Lagrange equations" Eq. \ref{mhpy} can be solved and plotted in a phase portrait mode that has been invoked previously by Pandit and Wortis to describe the phase equilibria in lattice models with surfaces and interfaces \cite{Wortis19982}.  The phase portrait method allows for an easy and physical visualization of the mean-field theories and has been successfully applied also to the case of the Poisson-Boltzmann case \cite{Gavish2016}. 

\begin{figure*}[t!]
\centering
\includegraphics[width=.8\textwidth]{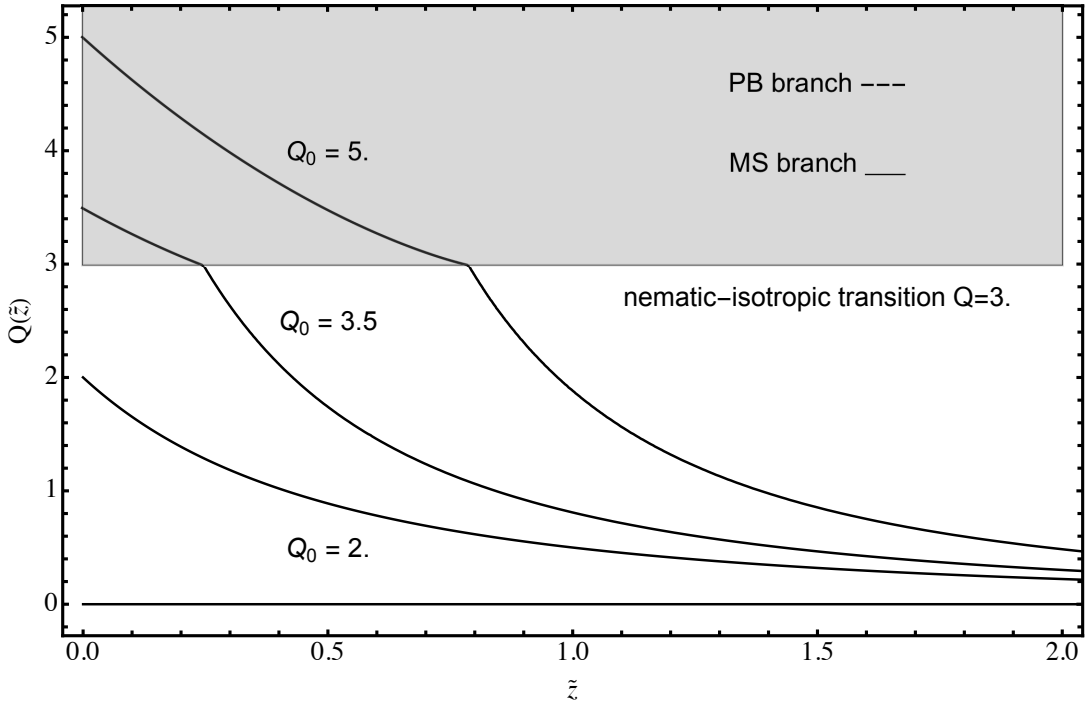}
\caption{The dimensionless counterion density profile $Q$ obtained from Eq. \ref{ghaek} as a function of the dimensionless separation from the bounding surface for different values of $Q_0$ as indicated on the figure. For $Q_0 \leq 3.$ the dimensionless density dependence on $\tilde z$ coincides with the PB branch. For $Q_0 > 3.$ it stays on the PB branch for $Q \leq Q_0$ but then exhibits a jump at the transition point and continues on the MS branch afterwards. The dimensionless density profile is continuous everywhere, but displays a discontinuity of the derivative at the wetting layer. In addition for $Q_0 > 3.$ the PB branch does not correspond to $Q_{PB}(\tilde z = 0) = Q_0$ at the surface but to a rescaled value such that $Q_{MS}(\tilde z = 0) = Q_0$, the difference stemming from the form of the boundary condition Eq. \ref{Hamilt5} on the MS and PB branches. \label{graph2}}
\end{figure*} 

One can now obtain the full implicit form of the solution of the mean-field equations  
by introducing two new variables
\begin{eqnarray}
P &=& \sqrt{u_{QQ}(0) \varepsilon} ~{\phi^*}'(z) \nonumber\\
Q &=&  u_{QQ}(0) \lambda_{(+)} ~ e^{-\phi^*(z) + \ln{J\big(\gamma(z)\big)}} = u_{QQ}(0) ~ c, \label{bdfka}
\end{eqnarray}
where $c  \equiv c_{+}$ is the counterion density, Eq. \ref{osmot}.  $Q$ can thus be interpreted as the dimensionless density of the rod-like counterions and $P$ is the dimensionless electrostatic field. It follows from the first integral Eq. \ref{ncfgjkws} that
\begin{eqnarray}
P = \pm \sqrt{ 2 Q + {\textstyle\frac{8}{27}} {\gamma^2(Q)}}, 
\label{Hamilt4}
\end{eqnarray}
where for a single charged surface the constant in the first integral can be obtained as vanishing, just as in the standard Gouy-Chapman case, meaning that the osmotic pressure of the system is zero. The solution of the problem is therefore completely specified by the dependence $P = P(Q)$, while $\gamma(Q)$, Eq. \ref{finalS}, becomes a solution of
\begin{eqnarray}
\gamma(Q)  = {\textstyle\frac{27}8}  Q ~\frac{\partial }{\partial \gamma(Q)}{\ln J\big(\gamma(Q)\big)}.
\end{eqnarray}
Numerical solutions of the above equations are presented in Fig. \ref{graph1} in the form of the dependence of the nematic order parameter ${\cal S}(Q)$ on the dimensionless rod-like counterion density, with the jump from zero to $0.313$ at the critical value $Q = 3.$, obviously corresponding to a first order isotropic-nematic transition.  The phase portrait of the system, $P = P(Q)$, presents the dependence of the dimensionless electrostatic field $P$ on the dimensionless density showing  the pure PB branchand the MS branch that separates from the PB branch at the nematic transition point. 

The surface charge density at the bounding surface sets the boundary value to
\begin{eqnarray}
P_0^2 = 2 Q_0 + {\textstyle\frac{8}{27}} {\gamma^2(Q_0)}  = {u_{QQ}(0)} \frac{\sigma^2}{\varepsilon},
\label{Hamilt5}
\end{eqnarray}
which follows directly from the Gauss boundary condition for the electrostatic field at a surface with surface charge density $\sigma$, {\sl i.e.} $\varepsilon  {\phi^*}'(z = 0) = \sigma$. The solution, $Q_0 = Q_0(\sigma)$, then implies that the $P(Q)$ curve starts at $P_0$ and then moves along the solution line to $P =0$, corresponding to the vanishing of the electrostatic field far from the surface. 

For $P_0$ large enough (see the $A$ intercept in Fig. \ref{graph1}) the solution first follows the MS branch in the ordered phase until it reaches the transition point. The system thus exhibits a finite thickness {\sl nematic wetting layer} close to the surface. After that the solution migrates to the PB branch, following it until the electrostatic field levels off to zero far away from the surface. On the other hand for $P_0$ smaller then a critical value  (see the $B$ intercept in Fig. \ref{graph1}) there is no nematic wetting and the system remains disordered along the whole $P(Q)$ solution, following the PB line as if the counterions are point-like. 

Clearly far away from the charged surface the system is disordered while in the proximity to the surface, where electrostatic attraction between the cations and the negatively charged surface increases their local concentration, it orders up, creating a surface wetting layer of the nematic phase. 

A note of caution is in order here: in the nematic wetting context the wetting layer and specifically its thickness is a result of the competition between the order parameter inhomogeneities, described by the order parameter gradient terms in the free energy, and the bulk free energy \cite{Sheng, Sluckin, Braun_1996}. The wetting layer and its properties in our case is related but does {\sl not coincide} with the the standard definition of nematic wetting. As already stated, we assumed that the spatial relaxation of both the nematic as well as the electrostatic fields is governed solely by the electrostatic component, so that in this respect the nematic field is subservient to the electrostatic field. A more complete but unfortunately also much more complicated theory would have to encompass both relaxations separately.

\begin{figure*}[t!]
\centering
\includegraphics[width=.8\textwidth]{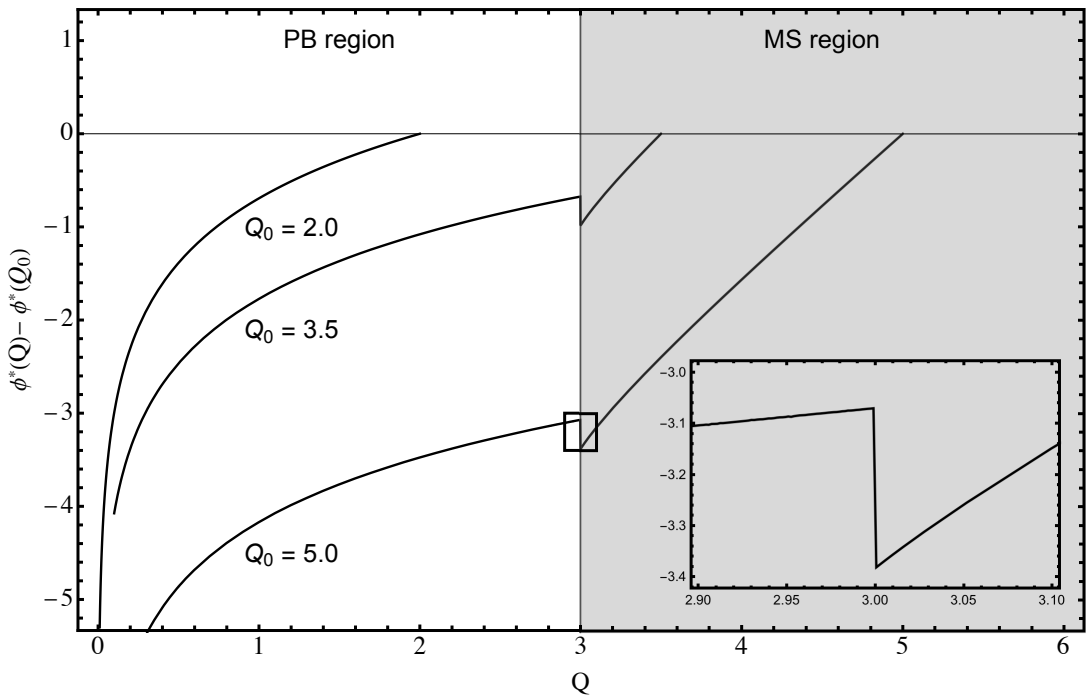}
\caption{Dimensionless mean-field electrostatic potential difference $\phi^*(Q) - \phi^*(Q_0)$ from Eq. \ref{ghaek1} as a function of the dimensionless distance from the surface $\tilde z$ for different values of $Q_0$, Eq. \ref{Hamilt5}, as indicated in the figure. For $Q_0 \leq 3.$ the electrostatric potential stays completely on the PB branch. For $Q_0 > 3.$ it stays on the PB branch for $Q \leq 3.$ but then exhibits a jump at the transition point and continues on the MS branch afterwards. The inset shows the details of the dependence for $Q_0 = 5.$. We refer to this jump at the transition point as the "Donnan potential difference". The value of  $\tilde z$ at this jump, that can be read off Fig. \ref{graph2}, also corresponds to the thickness of the nematic layer. In addition for $Q_0 > 3.$, $\phi^*(Q) - \phi^*(Q_0) = 0$ at the surface only for the MS branch while th ePB branch is suitably rescaled due to the form of the boundary condition Eq. \ref{Hamilt5}, on the MS and PB branches.  \label{graph3}}
\end{figure*} 

\subsection{Dimensionless counterion density and electrostatic potential}

The final dependence of the mean-field rod-like counterion density as well as the electrostatic potential on the dimensionless distance from the bounding surface, $\tilde z = q \sqrt{ \frac{2}{\varepsilon }} z$, is obtained implicitly from two equations. First by integrating the Poisson equation, Eq. \ref{MF22}, that can be rewritten in the form
\begin{eqnarray}
\tilde z = \int^{Q_0}_{Q(z)} \frac{~dQ}{Q} \left( \frac{\partial P(Q)}{\partial Q} \right),
\label{ghaek}
\end{eqnarray}
giving the dimensionless counterion density $Q$ as a function of $\tilde z$, and then obtaining the dependence of the dimensionless potential on the dimensionless density,  $\phi^*(Q)$, that can be derived from Eq. \ref{bdfka}, in the form
\begin{eqnarray}
\phi^*(Q) - \phi^*(Q_0) = \int^{Q_0}_{Q(z)} \frac{~dQ}{Q} \left( P(Q) \frac{\partial P(Q)}{\partial Q} \right),
\label{ghaek1}
\end{eqnarray}
so that $\phi^*(\tilde z)$ is obtained form $\phi^*(Q(\tilde z))$. The value of $Q_0$ is obtained from the boundary condition and Eq. \ref{Hamilt5} and  yields $Q_0 = Q_0(\sigma )$.  Clearly $Q_0$ can be either on the MS branch or the PB branch, as is clear from Fig. \ref{graph2}, and consequently the functional dependence on $\tilde z$ will depend on the $Q_0$, too. Note that the functional dependence $Q_0 = Q_0(\sigma )$ changes on the PB and the MS branch.

It is straightforward to obtain the limiting behavior for the dimensionless density from Eqs. \ref{ghaek}, \ref{ghaek1} which leads to
\begin{eqnarray}
\lim_{Q \gg 1} Q(\tilde z) = Q_0 ~e^{-\tilde z/a} \simeq Q_0 (1 - \frac{\tilde z}a)\qquad {\rm and} \qquad \lim_{Q \ll 1} Q(\tilde z) = \frac{1}{(\tilde z + \tilde z_0)^2},
\end{eqnarray}
where $a = \lim_{Q \gg 1} \frac{\partial P(Q)}{\partial Q}  = \sqrt{\textstyle 3/2}$ and $\tilde z_0^2 = 2/Q_0$. The former limit is valid for small separations, while the latter is valid  for large separations, coinciding with the standard counterion-only Gouy-Chapman result, as it should. 

Naively one would assume that the mean field electrostatic potential is proportional to the log of the density, just as in the standard GC case. However, the rod-like counterions also contain the orientational entropy as part of the mean field energy, see Eq. \ref{bdfka}, and thus the electrostatic potential is given rather by
\begin{eqnarray}
-\phi^*(Q)= \log{Q/\gamma(Q)\big)}.
\label{bdfka1}
\end{eqnarray}
While the spatial density profile is itself continuous with the derivative being discontinuous at the I-N transition, the electrostatic potential is discontinuous and displays a {\sl Donnan potential difference} at the transition point.  This Donnan potential difference, $\phi^*_D$, can be obtained straightforwardly from Eq. \ref{bdfka1} as
\begin{eqnarray}
\phi^*_D= \log{J\big(\gamma_{I-N}\big)} = \log{J\big({\textstyle\frac94} s_{I-N}\big)},
\label{bdfka2}
\end{eqnarray}
where $s_{I-N}$ is the jump of the orientational order parameter at the $I-N$ transition, and is therefore universal for all the electrostatic potential curves. It can be viewed as a Lagrange multiplier that ensures the electroneutrality of the system. The value of the Donnan potential difference across the phase boundary can be read off the graph, Fig. \ref{graph3}, as $0.313$, which equals exactly $ \log{\big(J({\textstyle 2.53})\big)}$ according to Eq. \ref{bdfka2}. 

\section{Discussion and conclusions}

While one can formulate the theory of inhomogeneous charged rod-like systems on different levels of approximations, we were specifically motivated to remain as close as possible to the Gouy-Chapman theory of point-like ions, the reason being that the Poisson-Boltzmann mean-field framework presents the foundation for the soft matter electrostatics and serves as a standard against which the new developments are usually compared with. Of course this necessarily implies also that the present theory shares at least the same weaknesses as those well known for the Poisson-Boltzmann theory \cite{Coulomb}.

The main feature of the theory presented is the two coupled mean-field equations which present generalizations of the standard Maier-Saupe and Poisson-Boltzmann equations. Their solution for a single charged surface with rod-like counterions leads to the existence of a nematic wetting layer, driven by the interplay of nematic and electrostatic interactions between charged rods and the bounding surface charges. The phase boundary at a finite distance from the surface is in addition characterized by an electrostatic potential jump corresponding to the nematic order parameter jump. While there are obvious similarities with the standard nematic-isotropic transition and the existence of nematic wetting, one needs to remember that in our case there is no chemical potential driving the density changes as well as no order parameter elastic energy driving the nematic wetting. The only driving force is electrostatics.

Among the possible and obvious generalizations of the present theory we should mention two explicitly. The first one is the tensorial nature of the interaction potential 
\begin{eqnarray}
u_{QQ}^{-1}({\bf x} -{\bf x}') \longrightarrow u_{iklm}^{-1}({\bf x} -{\bf x}'),
\label{xacbr432}
\end{eqnarray}
which would correspond to a more complicated liquid crystal elastic energy. The second one is the non-locality of the nematic interactions which implies the following form for the $u_{QQ}^{-1}({\bf x} -{\bf x}')$ interaction
\begin{eqnarray}
u_{QQ}^{-1}({\bf x} -{\bf x}') \longrightarrow u_{QQ}^{-1}(0)\left( 1 + \xi^2 \nabla^2\right) \delta({\bf x} -{\bf x}'),
\label{xacbr432}
\end{eqnarray}
or the corresponding tensorial expression consistent with Eq. \ref{xacbr432}. Above $\xi$ is the nematic order correlation length. The non-local form of the interaction potential  in its turn leads to a non-local form of the Maier-Saupe equation, Eq. \ref{MSgui1}, and one ends up with a system of two coupled non-linear differential equations. This would in its turn require also the introduction of the surface energy that would pin the surface value of the order parameter.  The no doubt complicated solutions would reduce to those studied above when the nematic correlation length $\xi$ is much smaller then the electrostatic correlation length, {\sl i.e.}, either the Gouy-Chapman length or the Debye length, depending on the composition of the system.

The other possible and obvious generalization would be to include the higher multipolar moments into the interaction energy, Eq. \ref{hamil1}, such as the quadrupolar electrostatic term. Formally this can be seen as leading to a modification in Eq. \ref{Maier} of the form
\begin{equation}
\Phi_{ij}({\bf x}) \longrightarrow \Phi_{ij}({\bf x}) + t \nabla_i\nabla_j \phi({\bf x}), 
\label{hdgkrynn}
\end{equation}
where $t$ is the strength of the quadrupolar moment of the charged rod. This generalization would treat the rod-rod electrostatic interactions more accurately, allowing for the existence of the Odijk effect (preferred perpendicular orientation of the rods) but would again imply a more complicated form of the Poisson-Boltzmann equation.

A variation on the geometry of the model could be pursued for a system confined between two charged surfaces with point-like co-ions. In this case one can either expect a surface nematic wetting transition or indeed a Fredericksz-type transition with a nematic phase between the surfaces and isotropic layers vicinal to the surfaces. These variations in the geometry setup would allow for interesting phenomena also in terms of the effective electrostatic interactions between the bounding surfaces that would no doubt deviate from the standard expectations.

\section{Acknowledgement}

This work was funded by the Key project \#12034019 of the National Science Foundation of China. I would also like to acknowledge the support of the School of physics, University of the Chinese Academy of Sciences, Beijing and of the Institute of physics, Chinese Academy of Sciences, Beijing. Finally I would like to thank D. Andelman, S. Buyukdagli and J. Everts for illuminating and helpful discussions and in particular M. Bier for valuable comments on an earlier version of this manuscript.

\section{Appendix}\label{Appendix}

Here we will give a short derivation of the field theoretic representation of the partition function as a functional integral over the scalar and tensor auxiliary fields based on the respective order parameters.  The grand canonical partition function  for the  system with Hamiltonian Eq. \ref{hamilt0} is defined standardly as
\begin{eqnarray}
{\Xi}[ {\bf r}_n, {\bf n}_n ]  = \sum_{N^+}\sum_{N_{(-)}} \frac{\lambda_{(+)}^{N^+} \lambda_{(-)}^{N^-}}{N^+!N^-!} ~\int\!\!\dots\!\!\int {\cal D}[{\bf x}_n] {\cal D}[{\bf n}_m] ~e^{-\beta {\cal H}[{\bf r}_n, {\bf n}_n]},
\label{part1}
\end{eqnarray}
where the integral over the orientational degrees of freedom, $[{\bf n}_m] $,  is only over the cationic species. Introducing the "decomposition of unit" in the form 
\begin{eqnarray}
1 &\equiv& \int {\cal D}[{\cal Q}_{ij}({\bf x})] ~\Pi_{{\bf x}} \delta({\cal Q}_{ij}({\bf x}) - \hat{\cal Q}_{ij}({\bf x})) \times \nonumber\\
&& \int {\cal D}[\rho({\bf x})] ~\Pi_{{\bf x}} \delta(\rho({\bf x}) - \hat{\rho}({\bf x})),
\end{eqnarray}
together with the functional integral representation with auxiliary potentials $\phi({\bf x})$ and $\Phi_{ij}({\bf x})$ for the functional delta functions,  one then remains with the following form of the partition function
\begin{eqnarray}
{\Xi}[ {\bf r}_n, {\bf n}_n ]   &=& \int {\cal D}[{\cal Q}_{ij}({\bf x})]{\cal D}[\Phi_{ij}({\bf x})] \times \nonumber\\
&& ~ \int {\cal D}[\rho({\bf x})]{\cal D}[\phi({\bf x})]~ e^{-\beta \tilde{\cal H}[{\cal Q}_{ij}({\bf x}), ~\rho({\bf x})]} \times \nonumber\\
&& \sum_{N^+}\sum_{N_{(-)}} \frac{\lambda_{(+)}^{N^+} \lambda_{(-)}^{N^-}}{N^+!N^-!} ~\int {\cal D}[{\bf x}_n] {\cal D}[{\bf n}_m]~ e^{ -\beta \tilde{\cal H}^*[{\bf x}_n, {\bf n}_m]}, \label{druga2}
\end{eqnarray}
where we introduced the field coupling part, ${\cal H}[{\cal Q}_{ij}({\bf x}), ~\rho({\bf x})]$, that stems from the two decompositions of unit as well as the interaction energy written with the collective coordinates, in the form
\begin{widetext}
\begin{eqnarray}
-\beta\tilde{\cal H}[{\cal Q}_{ij}({\bf x}), ~\rho({\bf x})] &=& -{\textstyle\frac12} \int\!\!\int_V d{\bf x} d{\bf x}' ~ {\cal Q}_{ij}({\bf x}) u_{QQ}({\bf x} -{\bf x}')  {\cal Q}_{ij}({\bf x}') - {\textstyle\frac12} \int\!\!\int_V d{\bf x} d{\bf x}' ~ {\rho}({\bf x})u_{\rho\rho}({\bf x} -{\bf x}') {\rho}({\bf x}') - \nonumber\\
&& ~~~~~~~~~ -i \int_V d{\bf x}~ {\cal Q}_{ij}({\bf x}) \Phi_{ij}({\bf x}) -i \int_V d{\bf x} ~\rho({\bf x})\phi({\bf x}),
\label{druga3}
\end{eqnarray}
\end{widetext}
while the configurational part in the internal space of the particles has the form
\begin{widetext}
\begin{eqnarray}
-\beta \tilde{\cal H}^*[{\bf r}_n, {\bf n}_n] &=& i \int_V\!\!\!d{\bf x} ~\hat{\cal Q}_{ij}({\bf x}) \Phi_{ij}({\bf x}) + i \int_V\!\!\!d{\bf x} ~\hat{\rho}({\bf x}) \phi({\bf x}) = %\nonumber\\
%&=&  
 i ~ \sum_{(+)} {\textstyle\frac32}\left({\bf n}_i {\bf n}_j - {\textstyle\frac13}\delta_{ij} \right)\Phi_{ij}({\bf x}_i)  +%\nonumber\\
%&& + 
~i  q\sum_{(+)} ~\phi({\bf x}_i) - i  \sum_{(-)} ~\phi({\bf x}_i), \nonumber\\
~
\end{eqnarray}
\end{widetext}
where we explicitly used the definitions of the microscopic scalar charge density and the microscopic tensor nematic order parameter density, Eqs. \ref{Qdef2} and \ref{Qdef1}. Since the ${\cal Q}_{ij}({\bf x})$ and $\rho({\bf x})$ functional integrals in Eq. \ref{druga2} are obviously Gaussian, these variables can be integrated out explicitly, yielding a pure field theoretical representation of the original configurational partition function Eq. \ref{part1} in terms of the tensor, $\Phi_{ij}({\bf x})$, and scalar, $\phi({\bf x})$, fields with
\begin{eqnarray}
{\Xi}[\Phi_{ij}({\bf x});  \phi({\bf x})]    &=& \int\!\!{\cal D}[\Phi_{ij}({\bf x})] \int\!\!{\cal D}[\phi({\bf x})]~ e^{-\beta{\underline{\cal H}}[\Phi_{ij}({\bf x});  \phi({\bf x})] }, \label{cgkfajbf}
\end{eqnarray}
where the effective field action is finally obtained in the form
\begin{widetext}
\begin{eqnarray}
-\beta {\underline{\cal H}}[ \Phi_{ij}({\bf x}); ~\phi({\bf x})] &=& ~-{\textstyle\frac12} \int\!\!\int_V d{\bf x} d{\bf x}' ~ {\Phi}_{ij}({\bf x}) u_{QQ}^{-1}({\bf x} -{\bf x}')  {\Phi}_{ij}({\bf x}') - {\textstyle\frac12} \int\!\!\int_V d{\bf x} d{\bf x}' ~ {\phi}({\bf x})u_{\rho\rho}^{-1}({\bf x} -{\bf x}') {\phi}({\bf x}') + \nonumber\\
&&  + ~\lambda_{(+)} \int_V d{\bf x} ~e^{ i\phi({\bf x}) +\ln{{\cal P}(\Phi_{ij}({\bf x}))}}+  \lambda_{(-)} \int_V d{\bf x} ~e^{-i\phi({\bf x})} -  
\nonumber\\
&& - {\textstyle\frac12} {\ln}~{\rm Det}{\left(u_{QQ}({\bf x} -{\bf x}')\right)} - {\textstyle\frac12} {\ln}~{\rm Det}{\left(u_{\rho\rho}({\bf x} -{\bf x}')\right)}.  
\label{fieldtyui}
\end{eqnarray}
\end{widetext}
Here we introduced the orientational partition function of a single particle 
\begin{eqnarray}
{\cal P}(\Phi_{ij}({\bf x})) \equiv \Big< e^{i {\textstyle\frac32}\left({\bf n}_i {\bf n}_j - {\textstyle\frac13}\delta_{ij} \right) \Phi_{ij}({\bf x})} \Big>_{\Omega}.
\label{Maier}
\end{eqnarray}
The orientational trace is defined in such a way that $<1>_{\Omega} = 1$.  The log of this expression is actually the orientational entropy of the rods. Were the field $\Phi_{ij}({\bf x})$ pure imaginary, the above distribution would correspond to the Bingham distribution, and the $ \Phi_{ij}({\bf x})$ field would be playing the role of the Bingham field.  

The formal identity of the configurational, Eq. \ref{part1}, and auxiliary field representations, Eq. \ref{cgkfajbf}, can be recapitulated as
\begin{eqnarray}
{\Xi}[ {\bf r}_n, {\bf n}_n ]  = {\Xi}[\Phi_{ij}({\bf x});  \phi({\bf x})] \label{partN}.
\end{eqnarray}
The steps leading to this identity are analogous to the case of Edwards transform for the Coulomb fluid partition function \cite{Coulomb}, except for the orientational Lebwohl-Lescher part that leads to a tensor order parameter and a tensor auxiliary field.

Notably, the two fluctuational ${\rm Tr} \log$ expressions at the end of the above equation pertain to the Casimir-type fluctuation terms \cite{Coulomb} and would be combined with the fluctuational determinant of the second order expansion of the above field theory. We will not delve into these details here.

%%%%%%%%%%%%%%%%%%%%  Bibtex %%%%%%%%%%%%%%%%%
%%%%%%%%%%%%%%%%%%%%%%%%%%%%%%%%%%%%%%%%%%%%%
%
%\bibliographystyle{ieeetr}
%\nocite{*}
\bibliography{paper-1.bib}

\end{document}